\documentclass[a4paper]{article} 
\newcommand{\de}{\mbox{d}} 
\newcommand{\e}{\mbox{e}} 
\newcommand{\NP}{{\em Nucl.~Phys.}} 
\newcommand{\ZP}{{\em Z.~Phys.}} 
\newcommand{\PR}{{\em Phys.~Rev.}} 
\newcommand{\PRL}{{\em Phys.~Rev.~Lett.}} 
\newcommand{\PL}{{\em Phys.~Lett.}} 
\newcommand{\nuc}[2]{{$^{#1}${#2}} } 
\usepackage{epsfig} 
\begin{document}

\begin{center}
%%   
%% 
%%            ------ TITLE ---
{\large\bf 
A microscopic investigation of the transition form factor 
\\ in the region of collective multipole 
excitations \\ of stable and unstable nuclei 
}\footnote{Work supported in part by 
Deutsche Forschungsgemeinschaft within the SFB 634 
and by the University of Athens under grant 70/4/3309.} 

\bigskip  
%%            ------ AUTHORS -----
{P.~Papakonstantinou}$^{a,b}$,
%\footnote{  
%{\em Email: }{\tt panagiota.papakonstantinou@physik.tu-darmstadt.de}}  
{E.~Mavrommatis}$^b$, 
{J.~Wambach}$^a$, 
{V.Yu.~Ponomarev$^{a,}$\footnote{Permanent address: 
JINR, Dubna, Russia.} 
} 

\bigskip 
%%           ------INSTITUTIONS ---
%% 
{\em 
$^a$Institut f\"ur Kernphysik, 
Technische Universit\"at Darmstadt,  
Schlossgartenstr.9, %\\ 
D-64289 
Darmstadt, Germany  
\\  
%%% 
$^b$Physics Department, Nuclear and Particle 
Physics Section, University of Athens, %\\ 
GR-15771 Athens, Greece 
} 

\end{center}

\begin{abstract} 
We have used a 
self-consistent Skyrme-Hartree-Fock plus Continuum-RPA
model to study 
the low-multipole 
response of stable and neutron/proton-rich Ni and Sn 
isotopes. 
We focus on 
the momentum-transfer dependence of the strength distribution, 
as it provides information 
on the structure of excited nuclear states 
and in particular on the 
variations of the transition form factor 
(TFF) with the energy. 
Our results show, among other things, that 
the TFF may show significant energy 
dependence in the region of the isoscalar giant 
monopole resonance  and that the TFF 
corresponding to the threshold strength in the case of 
neutron-rich nuclei is different compared to 
the one corresponding to the respective giant resonance. 
Perspectives are given for more detailed 
future investigations. 

\bigskip 

\end{abstract} 
%\end{frontmatter} 

% 
%-------------------------------------

\section{Introduction} 

The multipole response of nuclei at energies below the quasielastic 
regime is characterized by various collective excitations. 
Among these, giant resonances (GRs) of stable nuclei have been 
investigated for decades, both theoretically and experimentally 
\cite{Sp1991,GR00,COMEX1}. 
Macroscopically, they are interpreted as small-amplitude excitations 
corresponding to the propagation of zero sound in the nuclear medium. 
Collective modes 
of excitation at energies below GRs  
include surface vibrations and rotations of the nuclear 
body. 
Transverse modes of excitation, corresponding to oscillations 
of the current distribution, also exist - a well-known 
example being the twist mode \cite{Vin01} and perhaps 
the toroidal mode \cite{Rye2002}.  
At energies higher than GRs, overtones of GRs 
are examined \cite{ShS02,ShK03,GoU01,Cl01}.  
Despite the long-time effort devoted to the understanding 
of nuclear collective motion, several issues remain open; 
as far as the electric response of spherical 
stable nuclei is concerned, 
we may mention as examples the fine structure of GRs, 
the experimental energy-weighted sum rule (EWSR) of 
compression modes 
(isoscalar monopole and dipole resonances), 
the relation between the energy 
of the latter and the value of  
nuclear matter incompressibility, the identification 
of isovector quadrupole GRs and the nature 
of low-lying collective modes \cite{GR00,COMEX1}.  

Nowadays, it is feasible in addition 
to study experimentally the properties 
of unstable nuclei, in radioactive ion beam facilities. 
Access to their multipole response appears possible using 
inverse kinematics or double storage rings. 
The physics of nuclei far from stability 
is rich in new phenomena, 
owing to the presence of very weakly bound nucleons, the unusual 
ratio $N/Z$ (for given mass number) and the large difference 
between the proton and neutron Fermi energies. The structure of the 
particle continuum and the nature of low-lying strength 
in unstable nuclei is of particular importance, as 
they provide information on decay properties, polarizabilities, 
shell structure etc. 
At present, 
limited experimental information is available. 
The giant dipole resonance of neutron-rich Oxygen isotopes 
has been under study, as well as the dipole response of 
various light unstable nuclei (eg. neutron-rich He, Li, Be, C 
isotopes and proton-rich \nuc{8}{Be} and \nuc{13}{O})\cite{Au01}. 
Information is available on the low-lying quadrupole strength 
of various unstable nuclei \cite{RNT01}.  
Theory has been anticipating experiment to some extent, 
and has by now provided interesting predictions regarding 
the low-energy response and GRs of nuclei far 
from stability. 
For example, a considerable mixing of isoscalar and isovector 
transitions and an increased fragmentation of GRs 
are expected. A special feature of the response 
of very neutron-rich nuclei is the so-called 
threshold strength, namely 
the significant amount of 
isoscalar and non-collective  strength 
predicted to occur just above the neutron threshold. 
In the case of open-shell 
nuclei, the shell model \cite{Sag02} 
 and the Quasiparticle RPA (QRPA) 
\cite{KVG00,CoB01,Mat2001,KSG2002} have 
been applied. 
Relativistic RPA (RRPA) has been used as well \cite{VPR01,VPR2002}. 
Results on 
doubly magic nuclei have been obtained with standard RPA 
(with or without inclusion of higher-order configurations) 
or Continuum RPA (CRPA) methods; in particular, the 
self-consistent Skyrme-Hartree-Fock + Continuum RPA 
(SHF+CRPA) has been used for a systematic 
study of the 
response of nuclei far from stability by the group of 
I.~Hamamoto 
\cite{HSZ1996,HaS1996b,HSZ1997a,HSZ1997b,HSZ1999} 
and recently by our group \cite{PP2004,PWP2004}. 
For more information on the relevant 
theoretical and experimental activity, one can refer, 
for example, to the contributions in Ref.~\cite{COMEX1}.  

The RPA method describes satisfactorily the 
transition densities of collective states in stable nuclei. 
In the past the RPA has also been used to study transition 
densities and currents 
of individual excitations in the case of unstable nuclei 
and to compare with the behaviour of stable nuclei 
\cite{HSZ1996,HaS1996b,HSZ1997a,HSZ1997b,%
HSZ1999,PP2004,PWP2004,CLN1997a,CLN1997b,KST1997}. 
In this work we focus on the transition densities 
and form factors  
corresponding to electric excitations of 
stable and unstable nuclei, adopting an  
approach 
which allows a systematic study. 
Using the SHF+CRPA method, 
we will examine how the transition strength 
distribution of selected Sn and Ni isotopes, 
and for low multipolarities,   
varies with the momentum $q$ 
transferred to the system. 
For this we will consider an external field of the form 
$j_L(qr)$.  
The momentum dependence of 
the nuclear response can be studied experimentally using 
inelastic electron scattering. 
Further information accessible in 
inelastic electron scattering experiments 
concerns the transition current density, 
which has been the subject of a separate work \cite{PWP2004}. 

The purpose of the present pursuit is twofold.  
First of all, 
it is important to have a reliable microscopic description 
of the transition density characterizing 
a nuclear excitation - either of a stable or an unstable nucleus  - 
not only  to gain insight into the nature of the excitation,  
but also because theoretical transition densities enter 
the analysis of electron- or hadron-scattering  experiments 
aiming at identifying 
GRs. 
Theoretically, the transition density 
associated with a particular type 
of excitation approaches hydrodynamical behaviour 
in the case of very collective states, 
such as GRs, but 
in general it is expected to be energy dependent. 
Therefore, macroscopic descriptions may be inadequate 
in certain cases. 
Our approach will help us to pinpoint 
cases where the transition density varies significantly with 
excitation energy. Possible candidates are resonances with 
large fragmentation width.  
Second, we wish to contribute to the 
theoretical understanding, currently being formed,  
of the response of unstable nuclei, 
as a guidance to future experimental activity.  
Inevitably, some of the results obtained using 
conventional models  and interaction parametrizations 
(fitted to describe the properties of stable nuclei) 
 will be proven inaccurate 
by the future experimental research, but there is much to be 
learned in the process.

In Sec.~2 the quantities of interest are defined 
and our method of calculation is presented. Results 
for spherical, closed-shell Sn and Ni isotopes are presented 
and discussed in Sec.~3. Conclusions and perspectives are 
given in Sec.~4.

\section{Definitions and method of calculation} 
We consider the response of 
spherical, closed-(sub)shell nuclei 
to an external field of 
the type 
$${V}=\sum_{i=1}^AV_L(r_i,\tau_i)
Y_{L0}(\theta_i,\phi_i),$$ 
where the variable $\tau_i=p$ or $n$  
labels the isospin character -proton or neutron- 
of the $i$-th particle. 
For an isoscalar (IS) field, 
$V_L(r,p)=V_L(r,n)\equiv V_L(r)$ 
and for an isovector (IV) one 
$V_L(r,p)=-V_L(r,n)\equiv V_L(r)$. 
For $L=1$ we use an effective charge equal to  
$-\frac{N}{A}$ for protons and $\frac{Z}{A}$ 
for neutrons. 
In the following, the isospin label 
will be suppressed for the sake of simplicity. 

We set $$V_L(r)=[{4\pi (2L+1)}]^{1/2} j_L(qr),$$ 
where $j_L$ is a spherical Bessel function. 
In the long-wavelength limit $qR\rightarrow 0$ 
($R$ is the nuclear radius), 
we obtain the usual multipole operator  
of the form 
{ $r^K$, where $K=L$ for $L>0$ and $K=2$} 
for $L=0$. 
The transition density 
$\delta\rho_{L0} (\vec{r},E)$, 
characterizing 
the excited natural-parity state $|L0\rangle$ 
of energy $E$, %, 
is determined  
by its radial component  
$\delta\rho_L(r,E)$, where 
%%% 
\begin{equation} 
\delta\rho_{L0} (\vec{r},E) = (2L+1)^{-1/2} 
\delta\rho_L(r,E)Y_{L0}(\theta ,\phi )/r^2 
 . 
\end{equation} 
For IS~(IV) transitions, this is  
the sum~(difference) of the proton- and neutron-transition densities. 
The strength distribution 
$$S(E,q)=\sum_f|\langle 0 |V|f\rangle|^2 
\delta (E-E_f)$$  
(where $|f\rangle$ are the final states, excited by the 
$q-$dependent external field $V$ 
and $E_f$ their excitation energies) 
is related to the 
transition form factor $F_L(q^2,E)$, i.e. to the  
Fourier transform of $\delta\rho_{L0}(\vec{r},E)$. 
In particular, since we are 
dealing with continuous distributions, 
we write the strength in a small energy 
interval of width $\Delta E$ at energy $E$ as 
\begin{eqnarray} 
 S(E,q) %
&=& \frac{4\pi (2L+1)}{\Delta E} 
  |\int_0^{\infty} \de r \delta\rho_L(r,E) j_L(qr)|^2  
 \nonumber \\ 
&=& \frac{(2L+1)}{\Delta E} 
|\int\de^3r \delta\rho_{L0} (\vec{r},E) \e^{i\vec{q}\cdot\vec{r}}|^2 
\propto |F_L(q^2,E)|^2 . 
\end{eqnarray} 

As long as charge-current conservation 
is respected, 
the calculated $S(E,q)$ may be interpreted as 
the longitudinal response function 
entering the analysis of inelastic 
electron scattering within the Born approximation \cite{Ciofi1980}. 

Our aim is to search for variations  
of the transition density in the region of collective 
excitations. 
By looking at the strength distribution $S(E,q)$ 
corresponding to 
different values of 
momentum $q$, we will ``probe" 
the various momentum- (Fourier-) components of   
$\delta\rho_{L}$, according to Eq.~(2). 
In a scenario where 
the $\delta\rho_{L}$ (and $F_L$) 
turns out to be the same, within 
a proportionality factor $f(E)$, for all transitions 
making up a strength distribution $S(E,q)$, 
%$\delta\rho_L(r,E)=f(E)\delta\rho_L(r)$,  
the shape of the calculated $S(E,q)$, 
determined by $f^2(E)$, should not change with $q$. 
In realistic cases there will be non-trivial 
variations of the $\delta\rho_{L}$ 
with the energy, revealing the different microscopic structure 
of the various transitions. 

{ Following the standard RPA method, we consider particle-hole ($ph$) 
excitations, built on top of the Hartree-Fock (HF) 
nuclear ground state and subjected to the $ph$ 
residual interaction (HF+RPA method). In particular,} 
the quantities introduced above %of interest 
are calculated 
using a 
Skyrme - Hartree-Fock (SHF) plus Continuum - RPA (CRPA) model. 
For the HF ground-state, 
the numerical code of P.-G.~Reinhard \cite{ReXX} 
is used. 
The calculation of 
the response function (unperturbed HF, as well as RPA) 
%%%The RPA 
is formulated 
in coordinate space, 
as described in 
\cite{BeXX,BeT1975,VaS1981,vGi1983,RWH1988}. %%,BeT1975,VaS1981,RWH1988}. 
The radial part of the unperturbed $ph$ Green function, 
%with isospin character $\tau_z$ 
%and 
of multipolarity $L$ and specified
isospin character, is given by: 
\small 
\begin{equation} 
G_{L}^0                                %%% G_{\mu_1\mu_2 }^0 
(r,r';E) = 
\sum_{ph} \left\{  
\frac{ 
\langle p || O_L || h \rangle^{\ast}_r %%% C_{\mu_1 L}^{\ast} (r;ph) 
\langle p || O_L'|| h \rangle_{r'}     %%% C_{\mu_2 L } (r';ph) 
	  }{ 
      \varepsilon_{ph} -E}  % + i\eta } 
\pm %+ \pi_{\mu_1}\pi_{\mu_2} 
\frac{ 
\langle h || O_L'|| p \rangle^{\ast}_{r'} %%% C_{\mu_2  L}^{\ast} (r';hp) 
\langle h || O_L || p \rangle_r        %%% C_{\mu_1 L } (r;hp) 
	  }{ 
     \varepsilon_{ph} +E} % + i\eta }  
     \right\}  ,  
\label{Eghf} 
\end{equation} 
\normalsize 
where $O_L$ (or $O_L'$) is one of the operators 
%$\mu_i=1-6$ enumerate the operators 
$Y_{L}$, 
$ [Y_L\otimes (\nabla^2+{\nabla '}^2) ]_{L}$, 
$ [Y_{L \pm 1}\otimes (\vec{\nabla} - \vec{\nabla '})]_{L}$  
and $ [Y_{L \pm  1}\otimes (\vec{\nabla} + \vec{\nabla '})]_{L}  $.  
%In particular, the value 1 is reserved for the operator $Y_{L0}$. 
The sign of the second term depends 
on the symmetry properties 
of the operators $O_L$ and $O_L'$  
under parity and time-reversal 
transformations. 
Spin operators have been omitted in the present calculation. 
With $h$~% 
($
p 
$) 
we denote 
the quantum numbers of the   
 HF hole~(particle) state and  
 $\varepsilon_{ph}=\varepsilon_p-\varepsilon_h$ is the energy 
 of the unperturbed $ph$ excitation. 
The particle continuum is fully taken into account, 
as described in \cite{BeXX,vGi1983,ShB1975}. %% ,VaS1981}. 
A small but finite Im$E$ ensures that bound transitions 
acquire a finite width \cite{BeXX}. 
The RPA $ph$ Green function is given by 
the equation 
\begin{equation} 
G_L^{\mathrm{RPA}} = [ 1+G_L^0V_{\mathrm{res}}]^{-1} G_L^0 
 , 
\label{Egrpa} 
\end{equation} 
which is solved as a matrix equation in 
coordinate space, 
isospin character 
and 
operators $O_L$. 
The $ph$ {residual interaction} 
$V_{\mathrm{res}}$ is zero-range, 
of the Skyrme type, 
derived self-consistently from the Skyrme-HF energy 
functional \cite{vGi1983,Tsa1978}.   
%%%%\item 
%
From the Green function $G_L^{\mathrm{RPA}}$ 
for $O_L = O_L'=Y_L$ 
the strength distribution $S(E,q)$ is obtained as 
\begin{equation} 
S(E,q) = 
4 (2L+1) 
{\mathrm{Im}} 
\int j_L(qr)G_L^{\mathrm{RPA}}(r,r';E) %%% {11L}(r,r') 
j_L(qr') \de r \de r' 
 . 
\end{equation} 

{ Within the RPA, giant resonances (GRs) are generated 
as coherent $ph$ excitations between major $n\ell$ shells; 
for example, $\Delta N$=2 for the giant monopole and 
quadrupole resonances (GMR and GQR) and $\Delta N$=1 
for the isovector (IV) giant dipole resonance (GDR), where $N=2n+\ell$ 
is the energy quantum number of a shell. 
Therefore, a GR lies energetically in the neighbourhood 
of the $\Delta N \hbar\omega $ region, where the energy quantum 
$\hbar\omega\approx41A^{-1/3}$~MeV, within the 
simple harmonic-oscillator model. Its precise 
position relative to this value (lower or higher) is determined 
by the $ph$ residual interaction (attractive 
or repulsive). ``Overtones" of GRs are generated as 
$\Delta N'=\Delta N+2$ excitations. 
An example is the isoscalar (IS)  GDR, which 
lies energetically in the $3\hbar\omega$ region. 
The HF+RPA method can also describe low-lying 
%collective 
transitions with $\Delta N$=0 ($2^+$, $4^+$ etc), provided that the 
nucleus considered is not $n\ell -$closed.}

\section{Results and discussion}  

We have applied our method to six cases of spherical, 
closed-shell nuclei lying inside, close to or far away 
from the valley of stability. 
In particular, we have examined the 
isotopes \nuc{56,78,110}{Ni} and \nuc{100,120,132}{Sn}. 
\nuc{56}{Ni} is  
 the next heavier $Z=N$, doubly closed nucleus after 
\nuc{40}{Ca}. It is $\beta-$unstable, but lies very close 
to the stability line; being closed-shell, \nuc{56}{Ni} is 
the most stable Ni isotope where our 
model could be applied, since pairing is not included in the model.  
\nuc{78}{Ni} is a neutron-rich isotope, possibly doubly 
magic \cite{DGL00}.  
As was done before 
in Refs.~\cite{HSZ1996,PWP2004}, 
we have used the extremely neutron-rich 
configuration \nuc{110}{Ni} 
as an academic example of a closed-shell isotope 
in the vicinity of the neutron drip line. 
According to SHF results, the $N=82$ closure may still be 
valid in the Ni region \cite{HSZ1996}, although not conclusively. 
Lying on the proton-rich side of the nuclear 
chart and with a half life of the order of 
1s~\cite{Sch94,Fae95}, \nuc{100}{Sn} 
may be the heaviest $N=Z$ nucleus inside the 
proton drip line.  
\nuc{120}{Sn} is a stable, doubly magic Sn isotope, while   
\nuc{132}{Sn}  is a neutron-rich isotope with a half-life 
of 39.7s \cite{ABBW03} and among 
the most magic 
heavy nuclei, as no excited states of this nucleus 
have been detected below 4~MeV~\cite{BDZ01}. 

We have calculated the 
IS and IV monopole (ISM and IVM), IV dipole (IVD) 
and IS and IV quadrupole (ISQ and IVQ) 
response of the nuclei 
\nuc{56,78,110}{Ni} and \nuc{110,120,132}{Sn} 
using the RPA method described in the previous 
section, for 
$q=0.2,\, 0.4, \, 0.6, \, 
0.8, \, 1.0$~fm$^{-1}$. 
We have employed the Skyrme parametrization SkM*\cite{BQB82}, 
tailored to describe GRs of stable nuclei 
and used in previous studies of the response 
of exotic nuclei as well, 
eg. in Refs.~\cite{HSZ1996,HaS1996b,% 
HSZ1997a,HSZ1997b,HSZ1999,PP2004,PWP2004,CLN1997a,CLN1997b}. 
We have also used the Skyrme force MSk7 \cite{GTP2001}, 
whose parameters were determined by fitting the values of 
nuclear masses, calculated using the HF+BCS method, to the 
measured ones, for 1888 nuclei with various values of $|N-Z|/A$. 
The two forces have similar nuclear-matter properties, 
except for the effective mass $m^*$. The results obtained 
with the two Skyrme parametrizations agree qualitatively. 
Therefore, we only show results obtained with SkM*. 
  
\begin{figure} 
\begin{center}  
\epsfig{figure=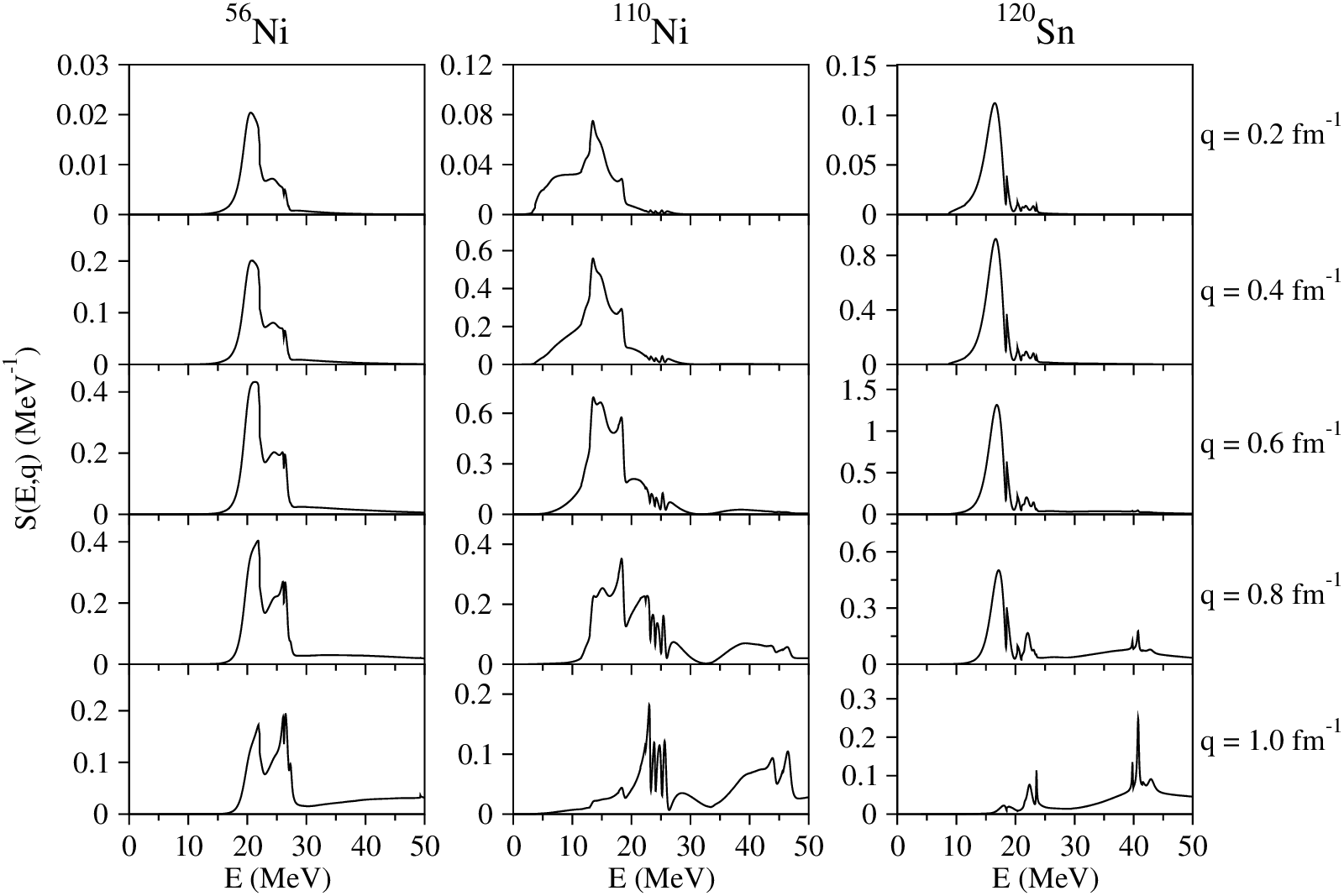,width = 11.5cm} 
\end{center} 
\caption{ISM strength distribution as a function of energy 
and momentum transfer. 
} 
\label{Fism} 
\end{figure}

\begin{figure} 
\begin{center}  
\epsfig{figure=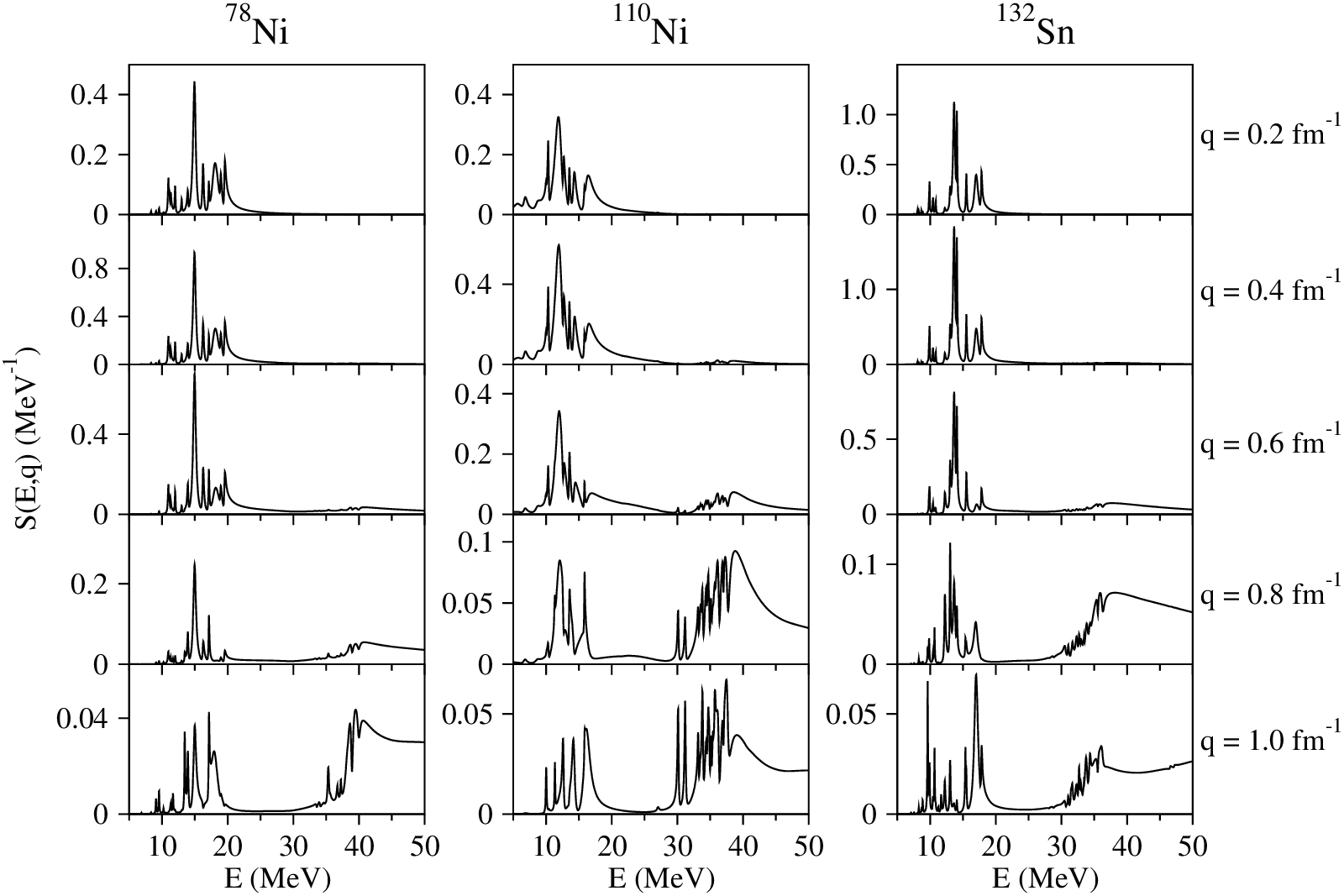,width = 11.5cm} 
\end{center} 
\caption{IVD strength distribution as a function of energy 
and momentum transfer. 
} 
\label{Fivd} 
\end{figure}

\begin{figure} 
\begin{center}  
\epsfig{figure=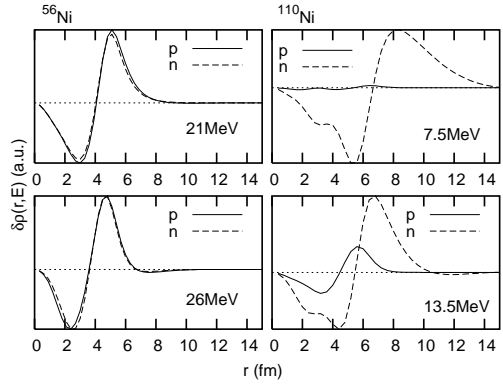,width = 5cm, angle = -90} 
\end{center} 
\caption{The transition density (in arbitrary units) 
for protons (full lines) and neutrons (dashed lines),  
corresponding to ISM transitions of the nuclei 
\nuc{56}{Ni} (left panel) and \nuc{110}{Ni} (right panel) 
at the indicated values of the energy. 
} 
\label{Fism1dr} 
\end{figure}

\begin{figure} 
\begin{center}  
\epsfig{figure=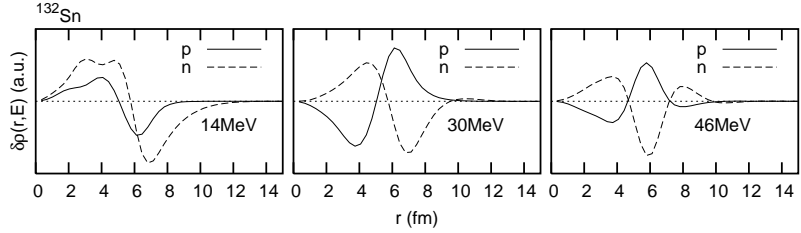,width = 3cm, angle = -90} 
\end{center} 
\caption{The transition density (in arbitrary units) 
for protons (full lines) and neutrons (dashed lines) 
corresponding to IVM transitions of the nucleus
\nuc{132}{Sn} 
at the indicated values of the energy. 
} 
\label{Fivm1dr} 
\end{figure}

Selected results are presented in Figs.~\ref{Fism}-\ref{Fivq}. 
In Figs.~\ref{Fism1dr}-\ref{Fivm1dr} the radial part 
of the proton- and neutron- transition density is plotted 
for some cases. We notice that,  
in all examined cases, as $q$ increases, the strength 
distribution is shifted to higher energies, which 
can be interpreted as the onset of  the quasielastic peak. 
The  continuum becomes increasingly important. 
Also, overtones of giant resonances 
become visible.  
For instance,  
in the ISM response, Fig.~\ref{Fism}, 
strength is 
shifted from the $2\hbar \omega$ region { (the IS GMR)}  
to the $4\hbar\omega$ region. 
In the IVD response, Fig.~\ref{Fivd},  
strength is found in the $3\hbar\omega$ region as 
$q$ increases. 
Excitations of single-particle character, 
with density oscillations 
taking place in the interior of the nucleus, 
give rise to this behaviour of the form factor - 
cf., for example, Fig.~\ref{Fivm1dr}, right panel, where the 
IVM transition density of \nuc{132}{Sn} at 46~MeV 
is shown. 
In the medium-heavy nucleus 
\nuc{56}{Ni} 
the shift takes place more slowly 
as a function of $q$ 
than in heavier nuclei, such as  
\nuc{120}{Sn}, because a narrower density 
distribution corresponds to a broader form factor.

Next we discuss our results in more detail. 

\subsection{Isoscalar monopole response 
and compression modes} 
In Fig.~\ref{Fism} the ISM response 
of \nuc{56}{Ni}, \nuc{110}{Ni} and \nuc{120}{Sn} 
is presented. 
{ Observing the values of the strength distribution 
of \nuc{120}{Sn} for the various values of $q$, we notice that} 
the form factor of the  weak peaks 
 on the high-energy tail of the GMR  %of  \nuc{120}{Sn} 
 has a weaker $q-$dependence, between $q=0.4-0.8$~fm$^{-1}$,  
compared to the GMR peak. 
The same holds for 
the other Sn isotopes and for \nuc{78}{Ni} (not shown). 
The GMR width of the medium-heavy nucleus \nuc{56}{Ni} 
is large,  compared to the GMR width of heavier nuclei. 
In Fig.~\ref{Fism} (left panel) 
two 
distinct energy regions can be recognized in the 
ISM strength distribution of \nuc{56}{Ni}:  
the region $P_<$ below 
$\mathcal{E}_0\approx 23$~{MeV}, 
and the region $P_>$ above 
$\mathcal{E}_0$. 
According to macroscopic models, 
the GMR is a uniform compression mode whose  
transition density has a node at the nuclear
surface. 
The node would then  
occur at radius $R\approx 1.2 A^{1/3}=4.6$~{fm} 
(or 4.3~fm, if we use our SHF result for the radius) 
in \nuc{56}{Ni}. 
Therefore, 
the transition density 
would show maximal overlap with the function 
$j_0(qr)$ (whose 1st root equals $\pi$)  
for 
$q=q_{01}=\pi /R=0.68$~(or 0.73)~{fm}$^{-1}$. 
It seems that the form factor 
in the region 
$P_<$ 
follows, at least approximately, this type of behaviour, 
since it reaches a maximum between  
0.6 and 0.8~{fm}$^{-1}$. 
The form factor in $P_>$ 
is maximized at a larger value of $q$ 
and therefore it does not correspond to such 
a picture. Indeed, as we observe in Fig.~\ref{Fism1dr}, 
the transition density in the regime $P_>$ has a node 
at a smaller radius than in $P_<$, 
by about 0.5~fm, and one more node at a larger radius. 

The fragmentation of the GMR is a typical feature 
of light and medium-heavy nuclei. 
It seems therefore important, for an accurate 
evaluation of the EWSR and the centroid energy of the GMR in 
those nuclei, to take into account the energy dependence 
of the transition density in the analysis of 
the relevant $\alpha-$scattering 
experiments  
 - see also the analysis in Ref.~\cite{KST2000}. 
The same may hold for the IS giant dipole resonance (GDR), 
which, in addition, 
appears to have large width even for nuclei 
as heavy as \nuc{208}{Pb}. 
{ Moreover, the amount and nature of the strength detected below the main 
IS GDR peak (which is located at 
approximately 23~MeV for heavy nuclei) 
has not been clarified. For a recent report 
on the issue } see eg. the contribution of 
U.~Garg in Ref.~\cite{COMEX1}. 
Given that the properties of the IS GMR and GDR are used 
for determining the value of nuclear matter incompressibility, 
a detailed examination may be recommended. We have not 
presented calculations of the IS GDR here, as our method lacks full 
self-consistency (due to the omission 
of spin-depended terms from the residual interaction) 
 and therefore our results would 
not be free of spurious components.

\subsection{Isoscalar quadrupole response}  

\begin{figure} 
\begin{center}  
\epsfig{figure=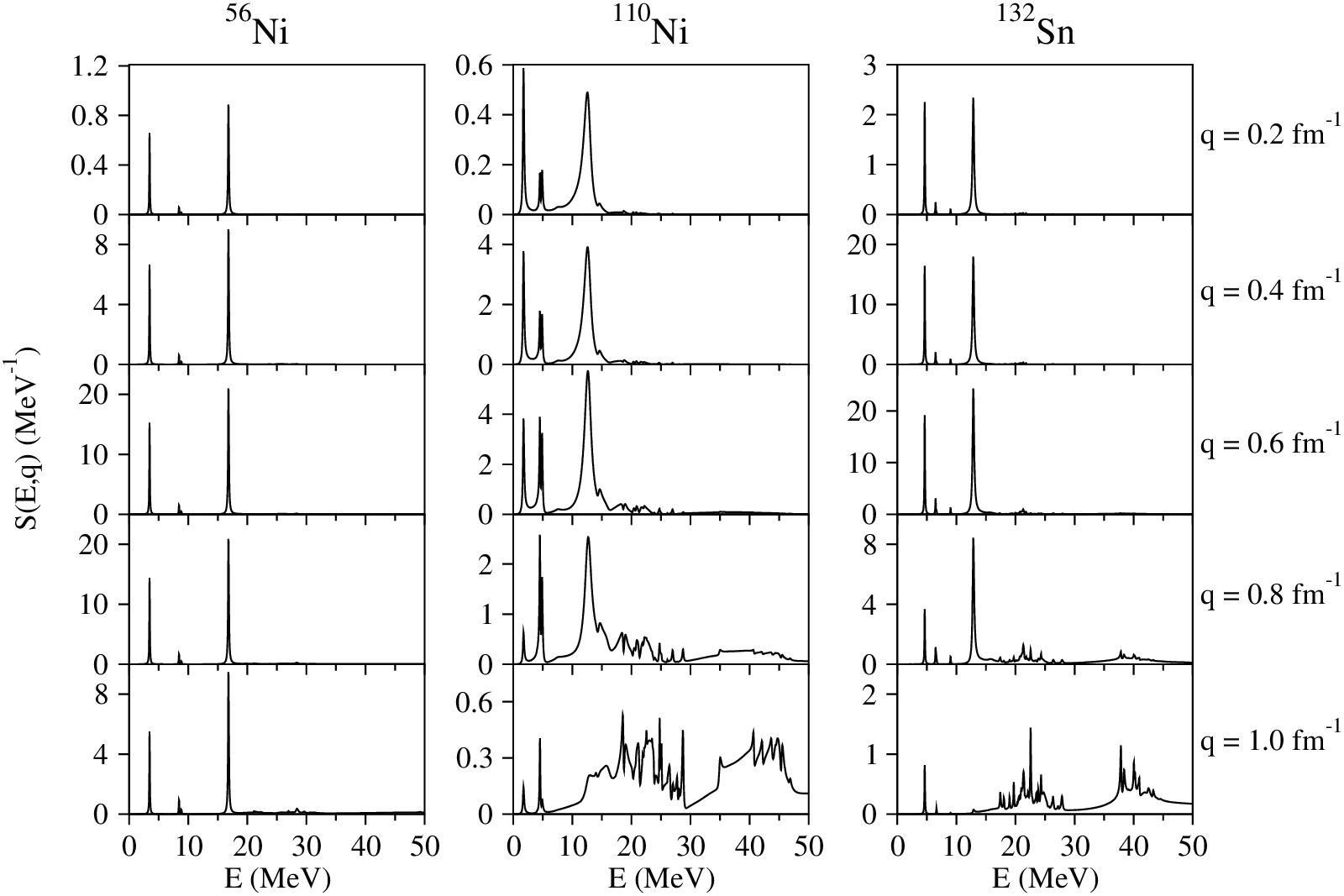,width = 11.5cm} 
\end{center} 
\caption{ISQ strength distribution as a function of energy 
and momentum transfer. 
} 
\label{Fisq1} 
\end{figure}

\begin{figure} 
\begin{center}  
\epsfig{figure=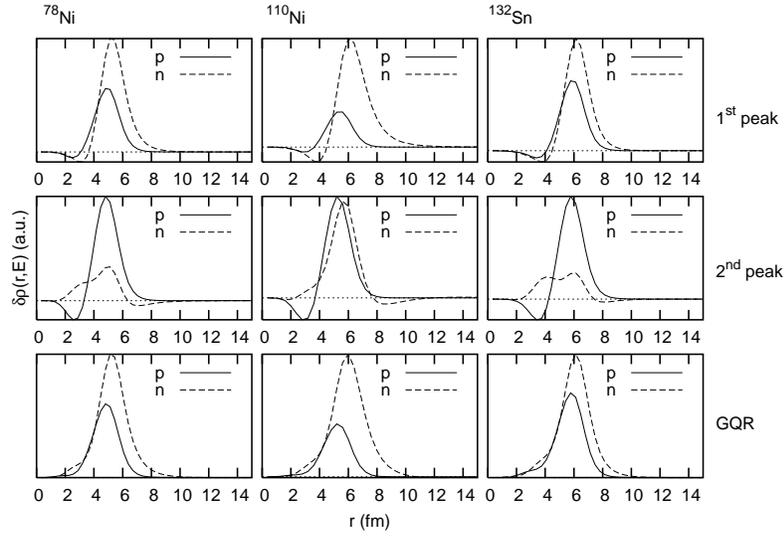,width = 7cm, angle = -90} 
\end{center} 
\caption{The transition density (in arbitrary units) 
for protons (full lines) and neutrons (dashed lines) 
corresponding to the first (collective) and second low-lying 
ISQ peaks and to the IS GQR, for the nuclei 
\nuc{78}{Ni}, \nuc{110}{Ni} and \nuc{132}{Sn}.  
} 
\label{Fisq1dr} 
\end{figure}

The two peaks in the ISQ strength distribution 
of the nuclei considered here 
are the low-lying collective $2^+$ transition (first peak) and 
the IS GQR (second peak). 
In the cases of  
\nuc{56}{Ni} (see Fig.~\ref{Fisq1}), 
\nuc{78}{Ni} (not shown)  and \nuc{100,120}{Sn} (not shown), 
the two peaks %namely the low-lying collective transition and the IS GQR, 
show similar behaviour as a function of $q$ 
up to   
$q\approx 0.8$~fm$^{-1}$. 
This is due to the fact that 
in these nuclei, the transition density of both states 
is peaked close to the surface. 
It should be noted, however, that the nature of the 
two peaks may be quite different, as the low-lying 
state is expected to be characterized by non-negligible 
vorticity \cite{PP2004,PWP2004}. 
We also find that the transition density corresponding to 
the low-lying state has a node inside the nucleus.  
In the case of 
\nuc{132}{Sn} (Fig.~\ref{Fisq1}),  
the form factor of the first peak 
does not follow the behaviour of the GQR form factor. 
In 
\nuc{110}{Ni} (Fig.~\ref{Fisq1}), 
there is a clear difference. 
The low-energy peak loses 
its strength faster than the GQR, { as $q$ increases,} 
behaving like the 
threshold strength of other multipolarities 
 - which we discuss in \S~\ref{Sthr}.  
Respective transition densities are plotted in Fig.~\ref{Fisq1dr}.  
In \nuc{110}{Ni}, a third peak occurs between the low-lying state and the GQR, 
also showing different behaviour. 
Its strength peaks at higher $q$ than the GQR and the low-lying-state 
strength. 
It is a non-collective state with a transition density 
localized in the 
interior of the nucleus. 
Such a peak occurs also in the case of 
\nuc{78}{Ni} (not shown) and \nuc{132}{Sn}. 
Corresponding transition densities 
are also plotted in Fig.~\ref{Fisq1dr}. 
They indicate that these secondary peaks are not of pure IS character. 

The experimental width and fragmentation 
of the IS GQR can be much larger than 
accounted for by first-order RPA calculations. 
%\cite{LAC01}. 
In order to reproduce 
the experimental width of the GQR one has to take into account 
higher-order configurations than $1p1h$ \cite{DNSW90,SW91}. 
Then one would be in 
a position to examine the energy dependence of the 
transition density in the region of the GQR.

\subsection{Isovector strength distributions}  

\begin{figure} 
\begin{center}  
\epsfig{figure=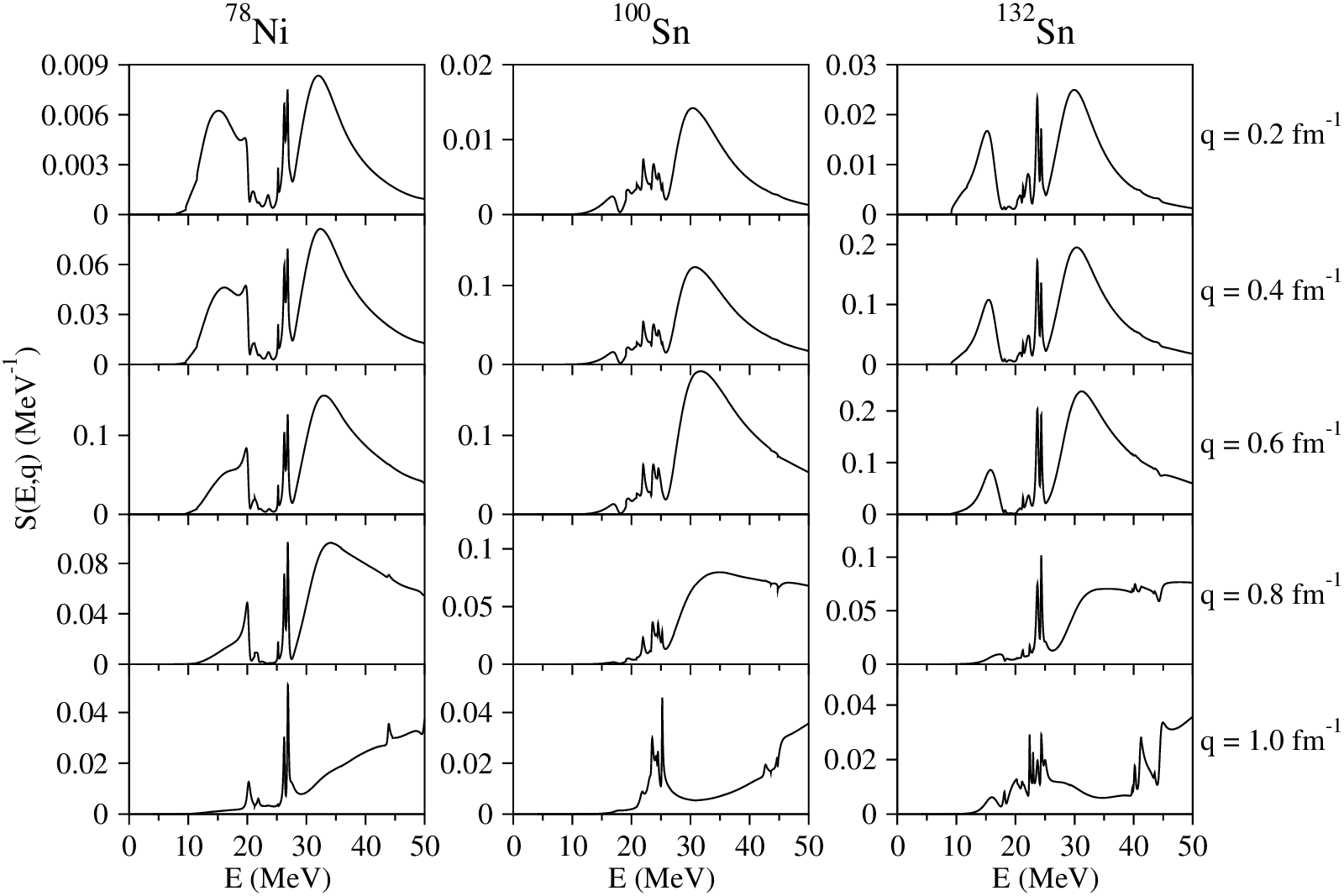,width = 11.5cm} 
\end{center} 
\caption{IVM strength distribution as a function of energy 
and momentum transfer. %Left: \nuc{78}{Ni}. 
} 
\label{Fivm} 
\end{figure}

\begin{figure} 
\begin{center}  
\epsfig{figure=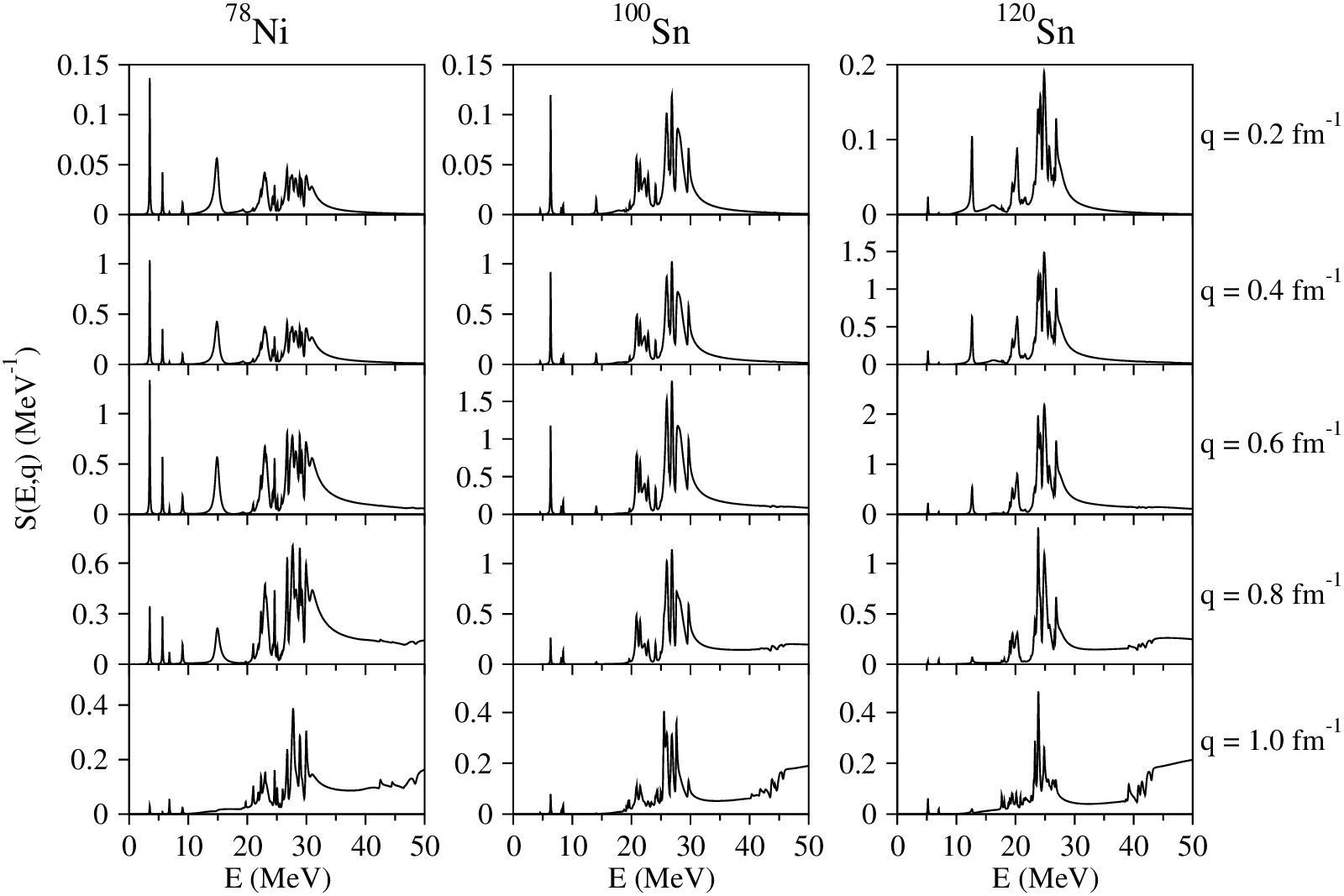,width = 11.5cm} 
\end{center} 
\caption{IVQ strength distribution as a function of energy 
and momentum transfer. 
} 
\label{Fivq} 
\end{figure}

As shown in Fig.~\ref{Fivm}, 
the IVM response of neutron-rich nuclei is 
dominated by two structures, namely the broad peak 
of the IV GMR { above 25~MeV}, 
higher than the IS GMR 
{due to the repulsive IV residual interaction}, and 
an amount of strength at the IS GMR region. The latter 
is of IS character (see Fig.~\ref{Fivm1dr}, left panel) 
and exemplifies the admixture between IS and IV transitions 
in neutron-rich nuclei. A similar situation is observed in 
the IVQ response (Fig.~\ref{Fivq}); 
{ most of the IVQ strength, making up the IV GQR, lies 
above 20~MeV, but an amount of strength remains in the 
IS GQR region.} In all nuclei, the IV GMR and 
GQR have a width of several MeV. 

No significant energy 
dependence of the transition density is observed in 
the region of the IV GMR and GQR for the 
low values of $q$ examined here. 
With the exception of the threshold strength 
in very-neutron-rich nuclei (see \S~\ref{Sthr}), 
the IVM strength distributions in Fig.~\ref{Fivm} 
for 
$q=0.2,0.4$ and 0.6~fm$^{-1}$ 
are similar to each other. 
The same seems to hold for the IVQ distributions 
(Fig.~\ref{Fivq}),  
in the region of the IV GQR, in spite of the 
rich fine structure of the latter. 
In the case of the IVD distributions in Fig.~\ref{Fivd} 
this holds to a lesser extent. 

The nature and systematics of the low-lying dipole strength 
is of particular importance in connection to 
astrophysical processes. A detailed examination 
of the IV as well as IS dipole strength distribution 
and also the corresponding vorticity strength distribution 
for various nuclei, 
including the momentum dependence of these distributions, 
using a fully self-consistent 
method, should be able to clarify - from a theoretical 
point of view - important issues such as the admixture of  
IS and IV transitions at low energies, the existence of 
toroidal modes and the energy dependence of the 
transition densities and currents in the region of the IS GDR.

\subsection{Threshold strength} 
\label{Sthr} 

\begin{figure} 
\begin{center}  
\epsfig{figure=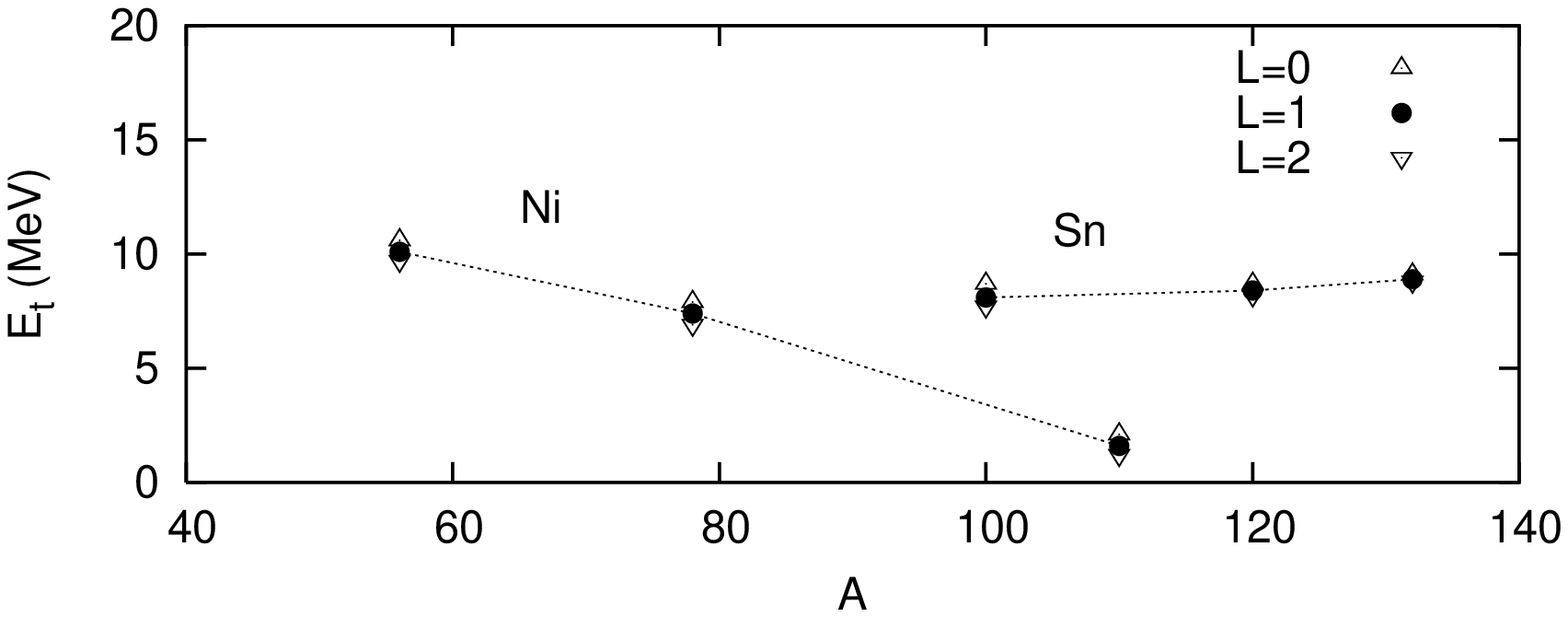,width = 4cm, angle = -90} 
\end{center} 
\caption{The calculated particle 
threshold energy $E_t$ plotted vs 
the mass number $A$, for the isotopes 
\nuc{56,78,110}{Ni} and \nuc{100,120,132}{Sn} 
and for $L=0,1,2$. Lines connecting 
isotopes of the same element are drawn to guide the eye. 
} 
\label{Fthresh} 
\end{figure}

As mentioned in the introduction, the response of 
very neutron-rich nuclei is characterized by 
the so-called 
threshold strength 
%, 
\cite{HSZ1996,HaS1996b,HSZ1997b,HSZ1999,CLN1997b}. 
{ The effect is expected when the neutron threshold 
lies much lower than the GR region - which begins 
typically at about 10~MeV. 
The values of the particle threshold energy $E_t$ for the isotopes 
\nuc{56,78,110}{Ni} and \nuc{100,120,132}{Sn} 
and for the multipolarilties considered here 
are presented in Fig.~\ref{Fthresh}. 
}  
According to our results, 
the threshold strength 
vanishes %rapidly 
as $q$ increases. 
This is seen clearly in the case of 
\nuc{110}{Ni}. 
As we observe in Fig.~\ref{Fism} (middle panel), 
the {ISM} threshold strength, 
{ appearing as a ``shoulder" 
at low energy,   
behaves differently than 
the peak at around 13.5~MeV; thus we have a way to discriminate 
between the ``threshold strength" and the main peak - 
which represents the GMR.}   
The same holds for the IVM response of \nuc{132}{Sn} 
(see  Fig.~\ref{Fivm}, right panel) and similarly for 
the IVD (see Fig.~\ref{Fivd}, middle panel). 
The distribution of loosely bound neutrons 
at large distances from 
the nuclear center results  
in form factors $F_L$ with maxima at  
low values of $q$, giving rise to this 
type of behaviour of the threshold strength. 
The effect of the excess neutrons 
and the origin of the threshold strength is demonstrated 
in Figs.~\ref{Fism1dr} (\nuc{110}{Ni}, 
compare right-top panel with the other panels) 
and \ref{Fivm1dr} 
(\nuc{132}{Sn}, left panel), where typical examples of 
transition densities corresponding to the threshold region 
are presented.  

In Ref.~\cite{HSZ1997b} the ISM response of the neutron-rich nucleus  
\nuc{60}{Ca} is examined - where a significant amount 
of threshold strength appears as well. 
It was found that in the region of the GMR 
the transition density compares rather well 
with the Tassie model prediction, 
whereas in the threshold region 
it is extended at large distances 
and it originates mostly from neutron excitations. 
A similar effect was predicted 
also in the case of the 
ISQ response of 
\nuc{28}{O} \cite{HaS1996b}. Our results are in 
concordance with these findings.

% 
%%%%%%%%%%%%% CONCLUSION %%%%%%%%%%%%%%%%%%%%%%%%%  
% 

\section{Conclusion and perspectives} 

In this work we have used the SHF+CRPA method 
to study the low-multipolarity response 
of selected Ni and Sn isotopes 
and to examine in particular 
variations of the transition density 
and form factor in the region of collective excitations. 
%distribution 
%varies with the momentum $q$ transferred to the 
%system. 
According to our results, the transition density 
may show considerable energy dependence in the region of the 
IS GMR. This should be taken into account in the 
analysis of relevant hadron scattering experiments. 
The form factor, corresponding to the threshold strength in 
very-neutron-rich nuclei, is narrower compared to the 
form factor of the respective giant resonance. 
This result, owing to the losely bound neutrons outside the core, 
 is independent of $L$. 
In the region of IV GMR and GQR resonances, 
no significant energy dependence 
is observed for $q$ values below 1.0~fm$^{-1}$. 
A detailed examination 
of the IS and IV dipole strength distribution, 
as well as the corresponding vorticity strength distribution, 
should be the subject of future work. Special care 
should be taken of the spurious center-of-mass motion.  
Important issues to be addressed include the admixture of 
IS and IV transitions at low energies, the existence of 
toroidal modes and the energy dependence of the 
transition densities and currents in the region of the IS GDR. 

%\ack 
%This work was supported in part by Deutsche Forschungsgemeinschaft 
%within the SFB 634 and by the University of Athens 
%under grant 70/4/3309. 

%% 
%%%%%%%%%%%%%%%% REFERENCES %%%%%%%%%%%%%%%%%%% 

%%%%%%%%%%%% BEGINNING OF FIGURES %%%%%%%%%%%%%%%%%%%%%%%

% 
%%%%%%%%%%%%%%%% ISM %%%%%%%%%%%%%%%%%%%%%%%%%%%%%%%%%%%% 
%%%%%%%%%%%%%%%%%%%%%%%%%%%%%%%%%%%%%%%%%%%%%%%%%%%%%%%%% 
% 
%%%%%%%%%%%%%%%% ISQ1 %%%%%%%%%%%%%%%%%%%%%%%%%%%%%%%%%%% 
%%%%%%%%%%%%%%%%%%%%%%%%%%%%%%%%%%%%%%%%%%%%%%%%%%%%%%%%% 
% 
%%%%%%%%%%%%%%%% IVM %%%%%%%%%%%%%%%%%%%%%%%%%%%%%%%%%%%% 
%%%%%%%%%%%%%%%%%%%%%%%%%%%%%%%%%%%%%%%%%%%%%%%%%%%%%%%%% 
% 
%%%%%%%%%%%%%%%% IVD %%%%%%%%%%%%%%%%%%%%%%%%%%%%%%%%%%%% 
%%%%%%%%%%%%%%%%%%%%%%%%%%%%%%%%%%%%%%%%%%%%%%%%%%%%%%%%% 
% 
% 
%%%%%%%%%%%%%%%% IVQ %%%%%%%%%%%%%%%%%%%%%%%%%%%%%%%%%%%% 
%%%%%%%%%%%%%%%%%%%%%%%%%%%%%%%%%%%%%%%%%%%%%%%%%%%%%%%%% 
% 
%\clearpage 
 
%%%%%%%%%%%%%%%% ISMdr %%%%%%%%%%%%%%%%%%%%%%%%%%%%%%%%%% 
%%%%%%%%%%%%%%%%%%%%%%%%%%%%%%%%%%%%%%%%%%%%%%%%%%%%%%%%% 
% 
%%%%%%%%%%%%%%%% ISQdr %%%%%%%%%%%%%%%%%%%%%%%%%%%%%%%%%% 
%%%%%%%%%%%%%%%%%%%%%%%%%%%%%%%%%%%%%%%%%%%%%%%%%%%%%%%%% 
% 
%%%%%%%%%%%%%%%% IVMdr %%%%%%%%%%%%%%%%%%%%%%%%%%%%%%%%%% 
% 
% 
%%%%%%%%%%%%%%%% IVMdr %%%%%%%%%%%%%%%%%%%%%%%%%%%%%%%%%% 
%%%%%%%%%%%%%%%%%%%%%%%%%%%%%%%%%%%%%%%%%%%%%%%%%%%%%%%%% 
% 
%%%%%%%%%%%%%%%%%% END OF FIGURES %%%%%%%%%%%%%%%%%%%%%%%


\begin{thebibliography}{10}
%

\bibitem{Sp1991} 
``Electric and Magnetic Giant Resonances in Nuclei", 
Ed.~Speth J, World Scientific 1991. 

%
\bibitem{GR00}
{Proc. Int. Conf. on Giant Resonances, {\rm Osaka, June 12-15
  2000}}, 
\newblock {\em \NP} A  {\bf 687} (2001).
%%CITATION = NUPHA,A687,;%% 

\bibitem{COMEX1}
{Proc. COMEX1, Int. Conf. on Collective Motion under Extreme
  Conditions, {\rm Paris, June 10-13 2003}},
\newblock {\NP} A {\bf 731} (2004).
%%CITATION = NUPHA,A731,;%% 


\bibitem{Vin01} 
Vinas X et~al 2001, 
{\PR} A {\bf 64} 055601; 
%%CITATION = PHRVA,A64,055601;%%  
Reitz B et~al 2002, 
{\PL} B {\bf 532} 179.   
%%CITATION = PHLTA,B532,179;%%  

\bibitem{Rye2002}
Ryezayeva N et~al 2002, 
\newblock {\PRL } {\bf 89} 
\newblock 272502.
%%CITATION = PRLTA,89,272502;%%  

\bibitem{ShS02} 
Shlomo S and Sanzhur A I 2002, 
{\PR} C {\bf 65} 044310. 
%%CITATION = PHRVA,C65,044310;%%  

\bibitem{ShK03} 
Shlomo S, Kolomietz V M and Agrawal B K 2003, 
{\PR} C {\bf 68} 064301. 
%%CITATION = PHRVA,C68,064301;%%  

\bibitem{GoU01} 
Gorelik M L and Urin M H 2001, 
{\PR} C {\bf 64} 047301. 
%%CITATION = PHRVA,C64,047301;%%  

\bibitem{Cl01} 
Clark H L, Lui Y-W and Youngblood D H 2001, 
{\PR} C {\bf 63} 031301(R). 
%%CITATION = PHRVA,C63,031301;%%  
 
\bibitem{Au01} 
Aumann T et~al 2001, 
{\NP} A {\bf 687} 103c and Refs. therein. 
%%CITATION = NUPHA,A687,103;%% 

\bibitem{RNT01} 
Raman S, Nestor C W and Tikkanen P2001, 
{\em At.~Data~Nucl.~Data~Tables} {\bf 78} 1. 
%%CITATION = ADNDA,78,1;%% 

\bibitem{Sag02} 
Sagawa H 2002, 
{\em Eur.~Phys.~J.} A {\bf 13} 87. 
%%CITATION = EPHJA,A13,87;%% 

\bibitem{KVG00} 
Khan E and Van Giai N 2000, 
\newblock {\em Phys. Lett.} B {\bf 66} 
\newblock 472. 
%%CITATION = PHLTA,B66,472%%    

\bibitem{CoB01} 
Colo G and Bortignon P F 2001,  
\newblock {\NP} A {\bf 696} 
427. 
%%CITATION = NUPHA,A696,427;%% 

\bibitem{Mat2001} 
Matsuo M 2001, {\NP} A {\bf 696} 371. 
%%CITATION = NUPHA,A696,371;%% 

\bibitem{KSG2002} 
Khan E, Sandulescu N, Grasso M and Van Giai N 2002, 
\newblock {\PR} C {\bf 66} 
\newblock 024309. 
%%CITATION = PHRVA,C66,024309;%% 

\bibitem{VPR01}
Vretenar D, Paar N, Ring P and Lalazissis G A 2001, 
\newblock {\NP} A { \bf 692} 
\newblock 496.  
%%CITATION = NUPHA,A692,496;%% 

\bibitem{VPR2002}
Vretenar D, Paar N, Ring P and Niksic T 2002,
\newblock {\PR} C { \bf 65} 
\newblock 021301(R).  
%%CITATION = PHRVA,C65,021301;%% 

\bibitem{HSZ1996}
Hamamoto I, Sagawa H and Zhang X Z 1996 
\newblock {\PR} C {\bf 53} 
\newblock 765.
%%CITATION = PHRVA,C53,765;%% 

\bibitem{HaS1996b}
Hamamoto I and Sagawa H 1996 
\newblock {\PR} C {\bf 54} 
\newblock 2369.
%%CITATION = PHRVA,C54,2369;%% 

\bibitem{HSZ1997a}
Hamamoto I, Sagawa H and Zhang X Z 1997 
\newblock {\PR} C  {\bf 55} 
\newblock 2361.
%%CITATION = PHRVA,C55,2361;%% 

\bibitem{HSZ1997b}
Hamamoto I, Sagawa H and Zhang X Z 1997 
\newblock {\PR} C {\bf 56} 
\newblock 3121.
%%CITATION = PHRVA,C56,3121;%% 

\bibitem{HSZ1999}
Hamamoto I, Sagawa H and Zhang X Z 1999  
{\NP} A {\bf 648} 
203. 
%%CITATION = NUPHA,A648,203;%% 

\bibitem{PP2004} 
Papakonstantinou P 2004, PhD Thesis, University of Athens.  

\bibitem{PWP2004} 
Papakonstantinou P, Wambach J, Mavrommatis E and Ponomarev V Yu 2004 
\newblock {\PL} B {\bf 604} 
\newblock 157. 
%%CITATION = PHLTA,B604,157;%%    

\bibitem{CLN1997a}
Catara F, Lanza E G, Nagarajan M A and Vitturi A 1997 
\newblock {\NP} A  {\bf 614} 
\newblock 86.
%%CITATION = NUPHA,A614,86;%% 

\bibitem{CLN1997b}
Catara F, Lanza E G, Nagarajan M A and Vitturi A 1997 
\newblock {\NP} A  {\bf 624} 
\newblock 449.
%%CITATION = NUPHA,A624,449;%% 

\bibitem{KST1997}
Kamerdzhiev S,  Speth J and Tertychny G 1997,  
\newblock {\NP} A {\bf 624} 
\newblock 328.
%%CITATION = NUPHA,A624,328;%% 

\bibitem{Ciofi1980} 
Ciofi degli Atti C 1980, 
\newblock {Prog.~Part.~Nucl.~Phys.} {\bf 3} 
163. 
%%CITATION = PPNPD,3,163;%% 

\bibitem{ReXX}
Reinhard P-G 1991, 
\newblock {``Skyrme-Hartree-Fock Model"}
\newblock {\em {in Computational Nuclear Physics I - Nuclear Structure}}, ed.
 Langanke K, Maruhn J E and Koonin S E (Springer, New York)
\newblock p.28.

\bibitem{BeXX}
Bertsch G 1991, 
\newblock ``The Random Phase Approximation for Collective Excitations", 
\newblock {\em ibid} 
\newblock p.75.

\bibitem{BeT1975}
Bertsch G F and Tsai S F 1975, 
\newblock {\em Phys. Rep. {\bf 18}}C 
\newblock 127.
%%CITATION = PRPLC,C18,127;%% 

\bibitem{VaS1981}
Van Giai N and Sagawa H 1981, 
\newblock {\NP} A {\bf 371} 
\newblock 1.
%%CITATION = NUPHA,A371,1;%% 

\bibitem{vGi1983}
Nguyen~Van Giai 1983  
\newblock in  {\em Nuclear Collective Dynamics} 
\newblock (World Scientific) p.356. 

\bibitem{RWH1988}
Ryckebusch J, Waroquier M, Heyde K, Moreau J and Ryckbosch D 1988 
\newblock {\NP} A {\bf 476} 
\newblock 237.
%\bibitem{Ryc1988}
%J.~Ryckebusch.
%\newblock Ph.D. thesis, University of Gent, 1988.
%%CITATION = NUPHA,A476,237;%% 

\bibitem{ShB1975}
Shlomo S and Bertsch G F 1975, 
\newblock {\NP} A {\bf 243} 
\newblock 507.
%%CITATION = NUPHA,A243,507;%% 

\bibitem{Tsa1978}
Tsai S F 1978, 
\newblock {\PR} C {\bf 17} 
\newblock 1862.
%%CITATION = PHRVA,C17,1862;%% 

\bibitem{DGL00} 
Daugas J M et~al. 2000  
{\PL} B {\bf 476} 
213.  
%%CITATION = PHLTA,B476,213;%% 

\bibitem{Sch94} 
Schneider R et~al 1994, {\ZP} A {\bf 348} 241;  
%%CITATION = ZEPYA,A348,241;%% 

\bibitem{Fae95} 
Faestermann T et~al. 1995, {GSI Annual Report}, p.~23. 

\bibitem{ABBW03} 
Audi G, Bersillon O, Blachot J and Wapstra A H 2003, 
{\NP} A {\bf 729} 3. 
%%CITATION = NUPHA,A729,3;%% 

\bibitem{BDZ01} 
Bhattacharyya P et~al. 2001 
{\PRL} {\bf 87} 
062502. 
%%CITATION = PRLTA,87,062502;%% 

\bibitem{BQB82} 
Bartel J, Quentin P, Brack M, Guet C and Hakansson H-B 1982 
{\NP} A {\bf 386} 
79. 
%%CITATION = NUPHA,A386,79;%% 

\bibitem{GTP2001} 
Goriely S, Tondeur F and Pearson J M 2001, 
{\em At.~Data~Nucl.~Data~Tables} {\bf 77} 311. 
%%CITATION = ADNDA,77,311;%% 

\bibitem{KST2000}
Kamerdzhiev S,  Speth J and Tertychny G 2000  
\newblock {\em Eur.~Phys.~J.} A {\bf 7} 
\newblock 483.
%%CITATION = EPHJA,A7,483;%% 

%\bibitem{LAC01} 
%Lacroix D, Ayik S and Chomaz P 2001, 
%{\PR} C {\bf 63} 064305 
%and Refs. therein. 

\bibitem{DNSW90} 
Dro\.zd\.z S, Nishizaki S, Speth J and Wambach J 1990, 
{\em Phys.~Rep.~{\bf 197}} 1. 
%%CITATION = PRPLC,197,1;%% 

\bibitem{SW91} 
Speth J and Wambach J 1991, ``Theory of Giant Resonances" 
in [1], p.1. 
\end{thebibliography}
\end{document}